# Dislocation-assisted linear complexion formation driven by segregation

Vladyslav Turlo [a], Timothy J. Rupert [a,b,*]

[a] Department of Mechanical and Aerospace Engineering, University of California, Irvine, CA 92697, USA

[b] Department of Chemical Engineering and Materials Science, University of California, Irvine, CA 92697, USA

* trupert@uci.edu

**Abstract**

Atomistic simulations are used to study linear complexion formation at dislocations in a body-centered cubic Fe-Ni alloy. Driven by Ni segregation, precipitation of the metastable B2-FeNi and stable $L1_0$-FeNi phases occurs along the compression side of edge dislocations. If the Ni segregation is not intense enough to ensure precipitate growth and coalescence along the dislocation lines, linear complexions in the form of stable nanoscale precipitate arrays are observed. Critical conditions such as global composition and temperature are defined for both linear complexion formation and dislocation-assisted precipitation.

Thermodynamically-stable grain boundary structures, known as *complexions*, have been intensively studied in recent years because of their significant impact on the mechanical [1, 2] and transport [3] properties of both coarse-grained and nanocrystalline materials [4-6], as well as thermal stability [7, 8]. Often formed due to dopant segregation, grain boundary complexions behave as equilibrium "phase-like" structures and can experience something similar to phase transitions, changing their type with respect to the system's temperature and composition [7, 9]. Nevertheless, such complexions exist only in the form of the planar structural defects, requiring the surrounding bulk to be in equilibrium.

Extending the concept of interfacial complexions to one-dimensional defects, or *linear complexions*, Kuzmina et al. [10] reported chemical and structural states confined at dislocations in a bcc Fe-9 at.% Mn alloy. These authors demonstrated that the structure and composition of these linear complexions correspond to an fcc Fe-Mn austenite structure. The phenomenon that Kuzmina et al. observed was heterogeneous second phase precipitation, caused by the local stress field and compositional inhomogeneity around the dislocations. Similar types of phenomena have been investigated both theoretically [11-15] and experimentally [16-19] in earlier studies as well. A theoretical investigation by Cahn [11] applied equilibrium thermodynamics to determine how the system's general characteristics such as Burger vector and supersaturation affect heterogeneous nucleation on dislocations. Recent theoretical works on this topic have used phase-field [14, 15] and kinetic Monte Carlo [12, 13] methods to investigate the kinetics of the nucleation, growth, and coalescence of a second phase. Nevertheless, the large number of empirical parameters that are needed limits the applicability of such models, providing only a qualitative agreement with experimental results. On the other hand, systematic experimental investigations have also been restricted due to the small size of precipitates formed along the dislocation lines and only a limited number of the binary and ternary alloy systems such as Al-Li [16], Al-Cu [17], Al-Ag-Cu [18], α-Fe-Nb-C[19] have been studied. These studies reported heterogeneous nucleation of precipitate arrays along the dislocation lines, which competes with homogeneous nucleation inside the grain to influence the final microstructure. The control of second-phase precipitation is extremely important for improving thermal stability (Zener pinning) [20, 21] and precipitation hardening [22, 23]. The linear complexions found by Kuzmina et al. [10] existed as arrays of stable nanoscale-size precipitates, which could potentially improve both the thermal stability and the strength of polycrystalline alloys. As a whole though, questions remain about how to control the process of second-phase precipitation and complexion formation on dislocations.

In this work, we report on atomistic simulations of segregation-induced formation of the intermetallic linear complexions confined at edge dislocations, using α-Fe doped with Ni. The metastable B2-FeNi intermetallic phase is first formed on dislocations and then partially transformed to the stable $L1_0$-FeNi intermetallic compound. It is important to note that this B2 phase is very small, meaning it would likely be invisible to electron diffraction techniques such as those used in Ref. [10]. This highlights the utility of atomistic simulations to identify nanoscale details of complexion transformations. Nucleation of the $L1_0$-FeNi phase on dislocations occurs for a wide range of the global compositions and annealing temperatures outside of the two-phase coexistence region on the equilibrium Fe-Ni phase diagram. For some of these compositions and temperatures, homogeneous nucleation is thermodynamically-restricted and the final microstructure is determined by dislocation-assisted precipitation, being limited by the amount of dopant segregation. For other cases, the solute segregation is strong enough to form large, bulk-like precipitates at the dislocations. Finally, if a balance is reached, true linear complexions in the form of nanoscale precipitate arrays can be formed along the dislocation lines.

Atomistic simulations were carried out using the Large-scale Atomic/Molecular Massively Parallel Simulator (LAMMPS) software package [24] and an embedded-atom method (EAM) interatomic potential for the Fe-Ni system [25]. The samples were equilibrated with a hybrid Monte Carlo (MC)/ Molecular Dynamics (MD) method, in which every MC step was followed by an MD equilibration over 100 integration steps of 1 fs each. MC steps were performed in the variance-constrained semi-grand canonical ensemble using a parallel algorithm developed by Sadigh et al. [26], which allows the system to reach the thermodynamic equilibrium state relatively quickly for a given composition and temperature. Semi-grand canonical ensemble is well-suited to explore phase transformations in multicomponent crystalline solids [26, 27], while it is also commonly used to investigate the formation of grain boundary complexions [7, 9]. In turn, MD steps were performed in an isothermal-isobaric ensemble at zero pressure, allowing the system to relax the local strain variations associated with solute segregation and phase transformations. A convergence criteria was used that requires the absolute value of the potential energy gradient over the final 20 ps to be less than 1 eV/ps. When this value is reached, the simulation cell has reached an equilibrium configuration and no major structural changes would be observed after this point. The equilibrium atomic structures were analyzed by polyhedral template matching (PTM) [28] and visualized with the OVITO software [29]. In all atomic snapshots, atoms were colored according to their crystal structure: violet – bcc Fe-Ni solid solution, green – $L1_0$-FeNi, red – B2-FeNi. The position of the dislocation core was determined by the dislocation extraction algorithm (DXA) [30] as implemented in OVITO.

A bcc Fe single crystal with one positive and one negative edge dislocation was first created, as shown in Figure 1(a). The edge dislocations were inserted by (1) removing one-half of the YZ atomic plane in the center of a sample and (2) equilibrating the system using molecular statics at zero pressure. Two types of simulation cells, "long" and "short" in terms of dimensions in the Z-direction, were considered. Both cells had dimensions of 23 nm and 24 nm in the X- and Y-directions, respectively. The dimensions of the Z-direction were equal to 75 nm and 7.5 nm for the long and short cells, respectively. The long samples equilibrated at 500 K were used to qualitatively characterize all aspects of the second-phase precipitation along the dislocation lines. Compositions in a range from 1-4 at.% Ni were considered. Because the long samples were computationally expensive, the short samples were used for more systematic studies over a broader range of global compositions (1-7 at.% Ni) and temperatures (300-800 K).

First, we investigated how a gradual increase in dopant concentration affects the structure of the long samples at 500 K. At global compositions below ~2 at.% Ni, the dopant atoms segregate to the compression side of the dislocation cores, forming equilibrium composition profiles similar to those reported in the literature [14]. Figure 1(b) shows local compositional variations for the 1.5 at.% Ni specimen. As shown at the black box in the bottom right corner of Figure 1(b), dopant segregation to the compression side of the dislocations is limited to a half-cylinder region that is roughly 2 nm in diameter, centered right below the dislocation core. The average composition of this half-cylinder region is used later to measure the segregation composition. Local composition increases much more quickly than the global composition. For example, an increase in the global composition from 1 to 1.5 at.% Ni leads to an increase in the local segregation composition from 3 to 11 at.% Ni. An increase in the global composition to 2 at.% Ni leads to precipitation at the compression side of the two dislocations, as shown in Figure 1(c). The precipitates have a near-equiatomic composition and correspond to a combination of the B2- and $L1_0$-FeNi intermetallic phases. In an experimental study of the fcc Al-Zr alloy [31], Nes reported a similar finding of metastable $L1_0$-$Al_3Zr$ phase formation on dislocations. The presence of metastable phases on dislocations is assumed to be caused by both the dislocation's stress field and the existence of low-energy coherent interfaces between the B2 phase and the bcc matrix. Figure 1(c) shows a more complex structure of precipitates with the two intermetallic phases coexisting on dislocations. Further increasing the global composition leads to coarsening of the precipitates and eventually their coalescence, as shown in Figure 1(d). To understand why precipitates grow only at the compression side of the dislocations, a slice of the XY plane bounded by the dashed black box in Figure 1(d) was analyzed in Figure 1(e)-(f). Chemical ordering is found in Figure 1(e) but there are a number of bent atomic layers and kinks, with the density of these defects generally increasing as one goes from the top to the bottom of the region. Figure 1(f) shows that there is a significant difference in the atomic volume between the bulk Fe phase and the intermetallic precipitate. Because the dislocation's stress field can compensate for the volume difference associated with the second-phase nucleation, one would expect the precipitates to grow along the dislocation lines.

Figure 2 represents the YZ and XZ projections of the systems discussed above, showing the intermetallic phase precipitation from two views. While the small FeNi precipitates stay only at the compression side of the dislocations and follow the dislocation lines (Figure 2(b)), the larger precipitates also grow away from the slip plane (Figure 2(c)). We do not find that the dislocation core is replaced by the interphase bcc-fcc boundary, as had been hypothesized in [32]. On the contrary, the dislocation persists and stays in its slip plane (see Figures 2(b)-(c)), but develops kinks following the [11-1] direction of the precipitate growth. At a global composition of 2 at.% Ni, this directional growth of the FeNi precipitates results in the formation of stable precipitate arrays along the dislocation lines (Figure 2(e)). This array is a true linear complexion, only stable at the one-dimensional dislocation defect, and analogous to nanoscale intergranular films in the grain boundary complexion literature. For higher global compositions, the formation of one large precipitate is observed. For example, an increase in the global composition to 2.5 at.% Ni leads to the formation of one precipitate (see Figure 2(f)) with a size in the [11-2] direction that is 2-3 times as large as was observed in the 2 at.% Ni sample. Consequently, precipitate interaction mechanisms such as merging, coalescence, or Oswald ripening may be involved in the formation of such a large final precipitate, if one were to track the kinetics of this formation.

In their experimental work, Kuzmina et al. [10] observed stable nanoscale precipitate arrays along the dislocation lines in an Fe-9 at.% Mn alloy, similar to those in Figure 2(e), and described these features as linear complexions. Figure 3 shows local compositional profiles through and along a similar complexion in our simulated Fe-Ni system. These profiles are similar to those shown in the fourth figure of the paper by Kwiatkowski da Silva et al. [32] for the experimentally investigated Fe-Mn system. Both experimental and simulated linear complexions demonstrate a repeating pattern of periodic elevation of the local solute content. Recently, Kwiatkowski da Silva et al. [33] demonstrated experimentally and by thermodynamic calculations that such repeating pattern along dislocation lines in the Fe-Mn system may be seeded by spinodal decomposition in the dislocation segregation zone. On the other hand, low-energy coherent interfaces among B2-FeNi and matrix phases may result in barrier-less nucleation of the intermetallic phase precipitates and their diffusion-controlled growth [14]. Finally, transition to the stable L10 phase seems to always happen inside of the B2 phase and have a structural nature, contrasting with what would be expected from a diffusion-controlled phase transformation. Unfortunately, the MD/MC methodology used in this work means that an exact transformation path cannot be tracked. Instead, we focus on exploring equilibrium states with segregation and intermetallic phase precipitation on dislocations.

The short sample geometries were used for a systematic study of segregation and intermetallic precipitation over a larger range of global compositions and temperatures. Figures 4(a)-(b) plot the results of these simulations in terms of temperature and local Ni composition. Blue markers represent the local segregation composition, as defined in Figure 1(b), and green and red markers represent the compositions of the $L1_0$ and B2 phases, respectively. The different marker types correspond to the different global compositions shown in Figure 4(c). This figure shows that the local segregation composition increases as global composition increases and temperature decreases. In Figure 4(a), we consider only the $L1_0$ phase and the segregation composition is measured for the samples where no such phase is found. This allows us for comparing the obtained results with the equilibrium phase diagram, where the metastable B2-FeNi phase is absent. If the dopant segregation at dislocation is strong enough to reach a near-equiatomic composition, the intermetallic phase precipitates with a composition corresponding to the $L1_0$-FeNi phase are formed (green markers in Figure 4(a)). In between these extremes, there are several interesting markers in the two-phase region of Figure 4(a), which correspond to the short samples with no $L1_0$ phase precipitation. In Figure 4(b), in which we consider only the B2 phase, most of these short samples demonstrates the formation of the Fe-rich B2 phase precipitates. Recalling that the long sample with 2 at.% Ni at 500 K demonstrated the formation of repeating precipitate arrays with both B2 and $L1_0$ phases (a true linear complexion), we hypothesized that the short sample geometry can restrict the nucleation of this repeating structure (the average size of precipitates along the [11-2] direction was ~10 nm, or more than the length of the short cell) and that the other samples in the two-phase region of Figure 4(a) would also form linear complexions. To test this hypothesis, two long samples with compositions of 4 and 5 at.% Ni were equilibrated at 600 K and one long sample with a composition of 1 at.% Ni was equilibrated at 400

K. The samples with 5 at.% Ni equilibrated at 600 K and 1 at.% Ni equilibrated at 400 K did in fact form a repeating array of nanoscale precipitates that were limited to the compression side of the dislocation line. On the other hand, the sample with 4 at.% Ni equilibrated at 600 K shows the same transition state with insufficient nucleation of the B2 phase, similar to the long sample with 1.5 at.% Ni equilibrated at 500 K that was shown in Figures 2(a) and (d). This means that this sample is likely very close to the boundary for complexion formation but there is slightly too little segregation.

Figure 4(c) summarizes our findings by updating the equilibrium phase diagram with new regions. Region (I) represents the two-phase region on the equilibrium phase diagram with intermetallic nucleation that just happens to occur at the dislocations. The $L1_0$ phase atomic fraction in precipitates increases with increasing Ni composition and decreasing temperature, reaching almost 50% in the sample with 7 at.% Ni equilibrated at 400 K. In region (I), the $L1_0$–FeNi phase precipitation is promoted by the local segregation to the dislocation line and formation of the metastable B2-FeNi compound, but the $L1_0$ precipitates could exist even without the further support of the dislocations' stress fields. In contrast, region (II) is outside of the bulk two-phase region and corresponds to dislocation-assisted precipitation. In this region, the strong solute segregation and local stress field leads to the phase transition near the dislocation lines, followed by the growth and coalescence of these intermetallic B2+$L1_0$ phase precipitates. This structure would not exist if the dislocations were removed. The thin region (III) shows an extension of the region (II), where stable precipitate arrays corresponding to true linear complexions are formed. Finally, region (IV) corresponds to the bcc Fe-Ni alloy with dopant segregation but no precipitation of any kind. While regions (I) and (IV) are well-known as part of the equilibrium phase diagram, regions (II) and (III) represent the conditions where the resultant microstructure is completely controlled by the presence of the dislocations. Although it is beyond the scope of this paper, we hypothesize that the stable nanoscale precipitate arrays corresponding to linear complexions in region (III) would restrict dislocation propagation significantly and consequently increase the strength of metallic alloys in a dramatic fashion. Since the defects also remain, one can imagine linear complexions as a way to stabilize dislocation networks within a microstructure.

**Acknowledgements**

This research was supported by U.S. Department of Energy, Office of Basic Energy Sciences, Materials Science and Engineering Division under Award No. DE-SC0014232.

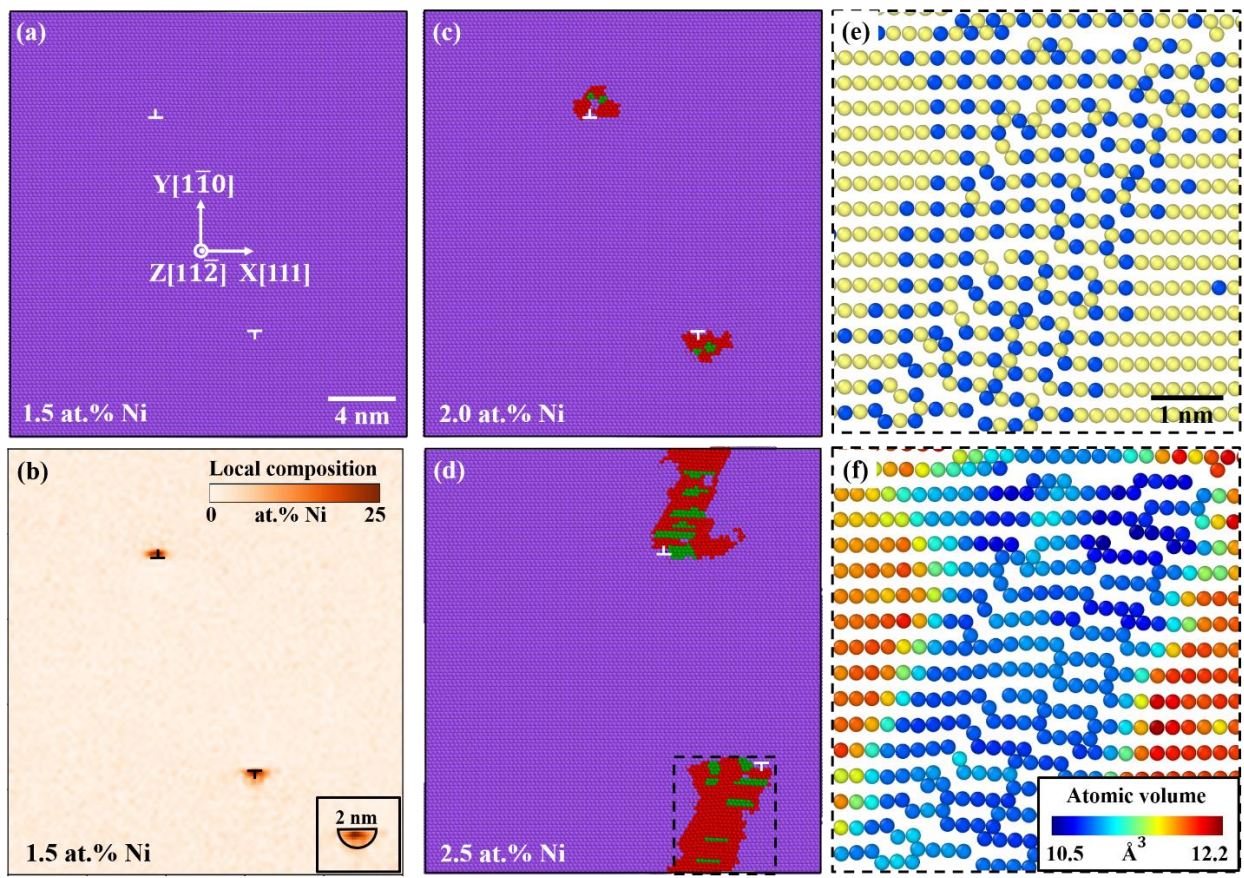

Figure 1. (a) View along the XY plane of the long simulation cell with two dislocations containing 1.5 at.% Ni and equilibrated at 500 K, as well as (b) the corresponding composition distribution. The small box at the right bottom corner of (b) shows the zone of intensive dopant segregation on the compression side of the lower dislocation. The B2+L1$_0$ phase precipitation is shown for the long samples with (c) 2.0 at.% Ni and (d) 2.5 at.% Ni at 500 K. Atoms in (a)-(d) are colored according to their alloy type: violet – bcc Fe-Ni solid solution, green – L1$_0$-FeNi, red – B2-FeNi. A slice along one atomic plane, limited to the dashed box in (d), is shown in (e,f). In these images, the atoms are colored according to (e) atom type (yellow – Fe, blue – Ni) and (f) atomic volume.

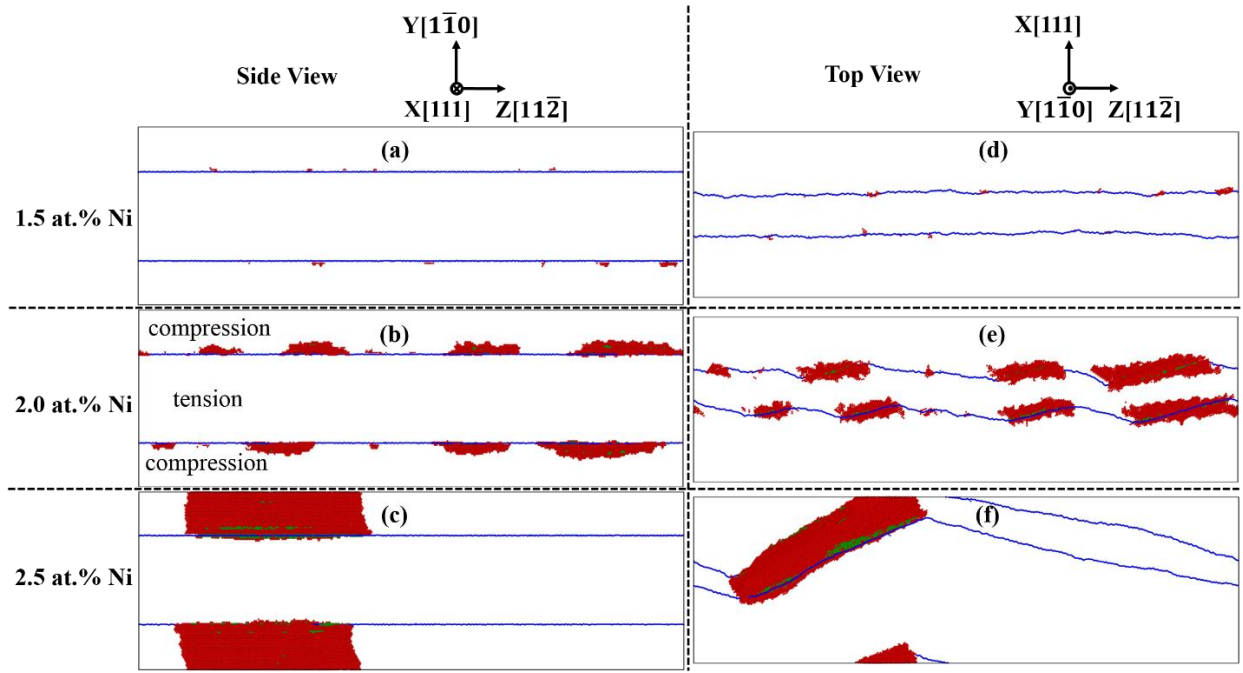

Figure 2. Views on the (a-c) YZ plane and (d-f) XZ plane for the long samples containing (a,d) 1.5 at.% Ni, (b,e) 2.0 at.% Ni, and (c,f) 2.5 at.% Ni, all equilibrated at 500 K. The dislocation lines appear blue, while the B2 and $L1_0$ intermetallic phase precipitates are shown in red and green, respectively. Atoms representing a bcc Fe-Ni solid solution are removed from the images.

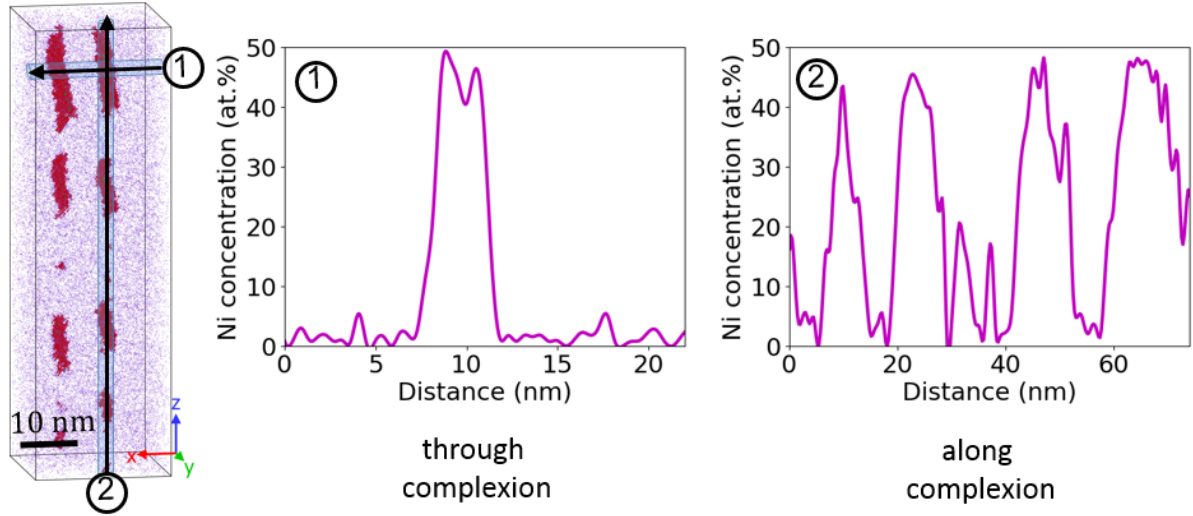

Figure 3. Visualization of linear complexions in the Fe-2 at.%Ni alloy at 500 K, along with compositional profiles (b) through and (c) along the dislocaton line.

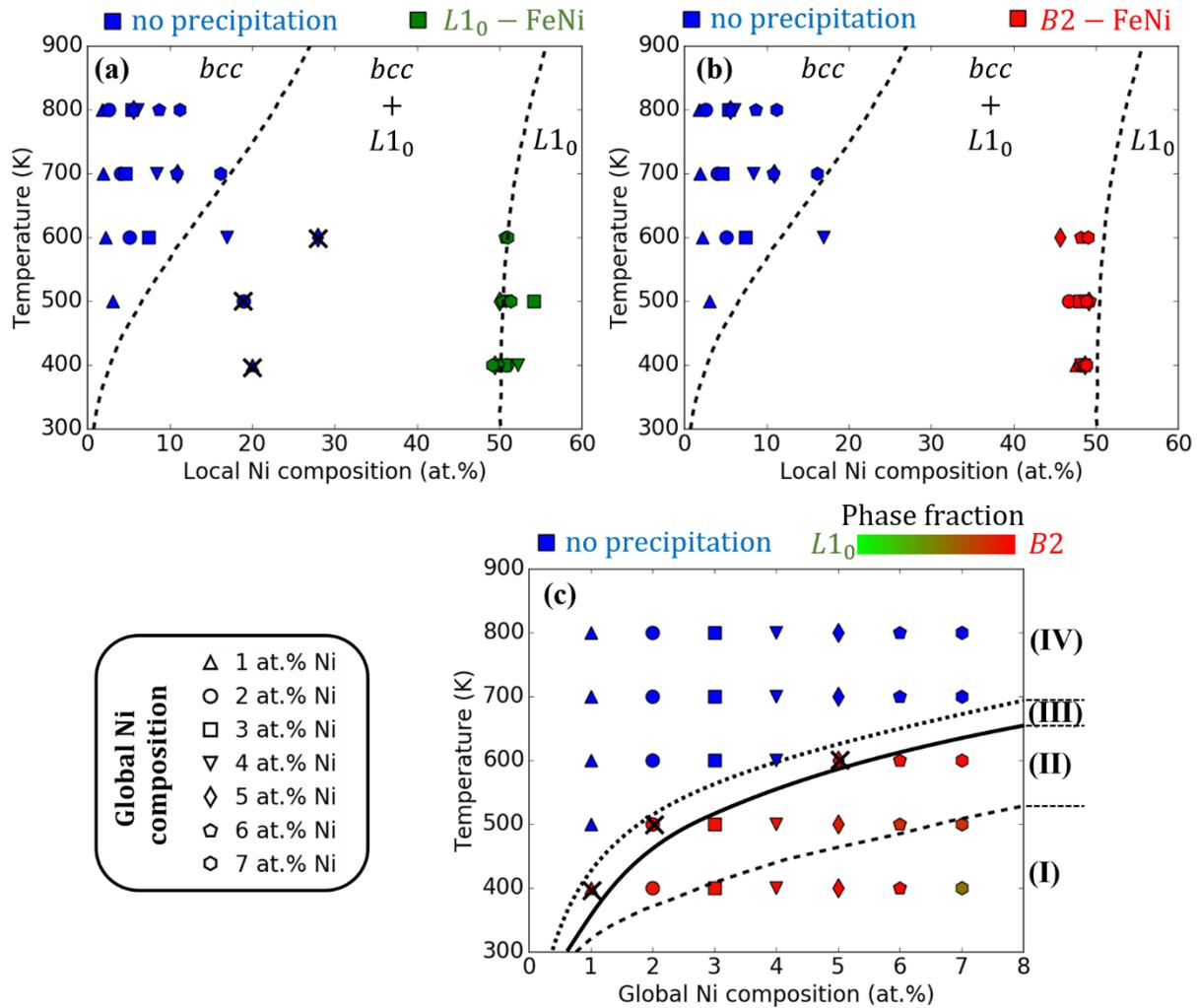

Figure 4. (a) Local composition–temperature plot for the short samples, where blue markers shows the local segregation composition and green markers show the $L1_0$ precipitate composition. (b) Local composition–temperature plot for the short samples, where blue markers shows the local segregation composition and red markers show the B2 precipitate composition. The markers labeled with X indicate the samples corresponding to the long samples with linear complexions. The marker shapes in (a) and (b) correspond to the different global compositions shown in (c). Dashed lines represent the equilibrium phase diagram curves for the Fe-Ni interatomic potential used in this study [25]. (c) The dislocation-affected phase diagram, in which four regions are found that correspond to (I) precipitation that happens to occur on dislocations, (II) dislocation-assisted precipitation on dislocations, (III) linear complexion formation along the dislocation lines in the form of nanoscale precipitate arrays, and (IV) dopant segregation with no precipitation. Limits of the regions (black curves) should be treated as approximations, created based on the observations of different atomistic configurations. Markers corresponding to precipitation on dislocations are colored according to atomic fractions of the B2 and $L1_0$ phases in precipitates.